\documentclass{article}
\usepackage[utf8]{inputenc}
\usepackage{authblk}
\usepackage[T1]{fontenc}
\usepackage{subeqnarray}
\usepackage{enumitem}
\usepackage{fullpage}
\usepackage{todonotes}
\usepackage{ bbold }
\usepackage{multirow}
\usepackage[title]{appendix}
\usepackage[cmex10,intlimits]{amsmath}
\usepackage{amssymb}
\usepackage{array}
\usepackage{mdwmath}
\usepackage{mdwtab}
\usepackage{siunitx}
\usepackage{cite}
\usepackage{graphicx}
\usepackage{subcaption}
\usepackage{upgreek}
\usepackage{hyperref}
\usepackage{amsthm}
\usepackage{amsmath}
\usepackage{color}
\usepackage{todonotes}
\usepackage{physics}
\usepackage{siunitx}
\AtBeginDocument{\RenewCommandCopy\qty\SI}
\usepackage[symbol]{footmisc}
\usepackage{mathtools}
\usepackage{float}
\usepackage{booktabs}

\graphicspath{{figures/}}

\title{Lisbon university deployed fiber Quantum Key Distribution with Post-Quantum bootstrapping}
\author[1,2]{Gonçalo Lobato Teixeira}
\author[3]{Manuel Goulão}
\author[1,2]{Emmanuel Zambrini Cruzeiro}
\author[1,2]{Preeti Yadav}

\affil[1]{Quantum Physics of Information (QPI) group, Instituto de Telecomunicações, Lisbon, 1049-001, Portugal}
\affil[2]{Quantum Information and Quantum Optics (QIQO) Laboratory, Instituto Superior Técnico, Lisbon, 1049-001, Portugal}
\affil[3]{INESC-ID, Instituto Superior T\'ecnico, University of Lisbon, Portugal}

\date{\today}

\begin{document}

\maketitle

\begin{abstract}
Towards the practical deployment of a stable quantum key distribution (QKD) system, we experimentally demonstrate a field-trial of a 1550 nm discrete-variable QKD system employing the efficient-BB84 protocol with an iPOGNAC-based transmitter over deployed optical fiber in a university campus network. The transmitter combines Sagnac-loop-based polarization modulation along with intensity modulation for the preparation of three polarization states with a single decoy level. We characterize the system performance at five channel losses, by reporting the measured quantum bit error rate (QBER) and secret key rate (SKR) for each case. We also demonstrate its stability over the fiber link corresponding to 16.68 dB of channel loss. This is achieved by reporting the fluctuations in the QBER and the mean photon numbers of the signal and decoy states over a period of 18 hours. The reported average QBERs are 0.24\% in the $X$ basis and 0.17\% in the $Y$ basis, despite not using any active polarization compensation mechanism. The average SKR is 8.2 kbps and never falls below 7.4 kbps throughout the long run. Moreover, the public discussion phase of our QKD system implements a post-quantum cryptography (PQC) based mechanism to achieve authentication. In contrast to the more widely explored approach of using PQC-based digital signatures for authentication, our implementation employs a PQC key encapsulation mechanism (KEM) to establish an initial shared symmetric key between the communicating parties. While this bootstrapping key is used to authenticate the classical communication in the very first QKD round, the subsequent rounds use QKD generated key to guarantee an information-theoretic security. Overall, the results demonstrate stable quantum-state preparation and QKD operation over an extended period of 18 hours in a realistic deployed-fiber environment, complemented by a one-time PQC KEM-based authentication.
\end{abstract}

\section{Introduction}
\label{sec:introduction}

Quantum key distribution (QKD) distributes shared symmetric keys between distant parties with security derived from the principles of quantum mechanics, rather than assumptions about the computational difficulty of mathematical problems \cite{bennett2014, gisin2002}. The resulting keys may subsequently be employed in symmetric cryptographic schemes, including the one-time pad for information-theoretic confidentiality or conventional encryption algorithms for computationally secure communications. An alternative approach to quantum-resistant key establishment is with post-quantum cryptography (PQC), which relies on computational problems conjectured to be safe even against quantum adversaries. Following the standardization of the first PQC algorithms by the National Institute of Standards and Technology (NIST) in 2024 with key encapsulation standards specified in FIPS 203 (ML-KEM), and digital signature standards in FIPS 204 (ML-DSA) and FIPS 205 (SLH-DSA), PQC offers a more practical approach towards quantum-resistant communications as compared to QKD. This is because QKD demands specialized optical hardware, dedicated quantum channel and authenticated classical communication. On the other hand, while PQC can be integrated into existing network infrastructures, its security guarantees remain dependent on computational hardness assumptions. These complementary properties motivate the investigation of hybrid architectures combining QKD with conventional and/or PQC mechanisms, which is often approached via PQC based authentication and/or hybrid key generation.

Among the different QKD protocols, the prepare-and-measure BB84 protocol remains the most extensively studied and experimentally implemented approaches for QKD \cite{bennett2014}. Practical implementations commonly employ phase-randomized weak coherent pulses (WCP) instead of ideal single-photon sources, introducing multi-photon emissions that can be exploited through photon-number-splitting (PNS) attacks. The decoy-state method addresses this vulnerability by randomly varying the mean photon number of the transmitted pulses, allowing the legitimate parties to estimate the contribution of single-photon events to the final secret key \cite{hwang2003, lo2005decoy, wang2005, ma2005practical}. Over the years, several developments have sought to further simplify the BB84 protocol, reducing experimental complexity while maintaining or improving its performance. In contrast to the original BB84 protocol, which employs four quantum states $\{|0\rangle, |1\rangle, |+\rangle, |-\rangle \}$ drawn from two mutually unbiased bases, Tamaki et al. established a security framework based on the use of only three quantum states \cite{tamaki2013}. Subsequently, Rusca et al. extended the security analysis to implementations employing a reduced set of measurements at the receiver, incorporating finite-key effects in what is known as the ``simplified-BB84'' protocol \cite{rusca2018security, grunenfelder2018, rusca2018}. Together, these contributions laid the foundation for simplified-BB84, enabling reduced complexity in both quantum-state preparation and detection while retaining rigorous security guarantees in the finite-key regime.

Beyond protocol design-based simplifications, the practical deployment of QKD requires stable quantum state preparation and reliable operation under environmental fluctuations. This is particularly relevant for polarization-encoded protocols, where imperfect state preparation and time-dependent polarization transformations can contribute to the quantum bit error rate (QBER), consequently reducing the achievable secret key rate (SKR). Therefore, conventional polarization-encoding implementations typically require careful calibration or active stabilization to maintain high state-preparation fidelity over extended periods. In this context, the polarization-encoding scheme based on a Sagnac loop, known as iPOGNAC, offers an attractive solution for stable and compact QKD transmitters \cite{agnesi2019, agnesi2020}. The iPOGNAC encoder solves this structurally using a self-compensating Sagnac loop, where controlled phase modulation is applied to two counter-propagating optical components (clockwise and anti-clockwise) in the same interferometric loop with a relative delay between them. Since both components traverse the same physical path, common-mode drifts cancel, and the device is calibration-free and stable without feedback, reaching 0.2\% QBER over 24 hours, as demonstrated in previous works \cite{avesani2020}.

The transition from controlled laboratory experiments to deployed optical-fiber operation introduces additional challenges associated with channel attenuation, connector and splice losses, environmental perturbations, and long-term polarization drift. Consequently, field-trial demonstrations are essential for evaluating the stability and practical performance of QKD systems under realistic operating conditions. At telecommunications wavelengths around 1550 nm, superconducting nanowire single-photon detectors (SNSPDs) offer high detection efficiencies, low dark-count rates, and excellent timing resolution, making them especially suitable for photon detection over lossy optical channels \cite{hao2024, fleming2025}. However, detector saturation, dead-time limitations, and polarization-dependent detection efficiencies must also be considered, as these can constrain the achievable key rate as well as introduce detection-efficiency mismatches and the associated side channel attacks \cite{marcomini2025}. The combined optimization of state preparation, optical transmission, and single-photon detection is therefore an important consideration in the practical implementation of stable and secure QKD systems. A notable work by Ma et al. demonstrates the long-term stability of a QKD scheme with a Sagnac loop-based intensity and polarization modulation over 75 km of fiber spool for a duration of 48 hours without any active polarization compensation \cite{ma2021simple}. More recently, a high-speed and low-error 3-state 1-decoy efficient-BB84 scheme \cite{grunenfelder2018} (without a simplified setup for the receiver) with the iPOGNAC transmitter is demonstrated \cite{bolanos2026ghz}. 

In this work, we experimentally demonstrate a polarization-encoded efficient-BB84 QKD system employing an iPOGNAC-based transmitter over deployed optical fiber within the Lisbon University network. The system implements the 3-state 1-decoy BB84 protocol using commercially available optical and electronic components, together with SNSPDs at the receiver. We characterize its performance in terms of the QBER and SKR under different channel losses, reaching losses of approximately 25 dB. Furthermore, we investigate the long-term stability of the system through continuous operation over a 18 hour period, without any active polarization compensation. During this experiment, we monitor the signal and decoy state intensities, as well as the evolution of the QBER and SKR, to assess the system's stability under realistic field conditions. Moreover, to address the scalability of QKD networks without relying on pre-shared symmetric keys for authenticating the classical communication, we integrate a PQC KEM to bootstrap a shared secret for the very first QKD round. Together, these results provide an experimental assessment of efficient-BB84 QKD operation over deployed optical fiber in the university of Lisbon campus, while demonstrating a practical approach for a one-time PQC secure authentication with classical cryptographic infrastructures.

\section{Efficient-BB84 QKD protocol with iPOGNAC emitter}
\label{sec:efficient_Pognac}

\begin{figure}[ht]
    \centering
    \includegraphics[width=0.85\textwidth]{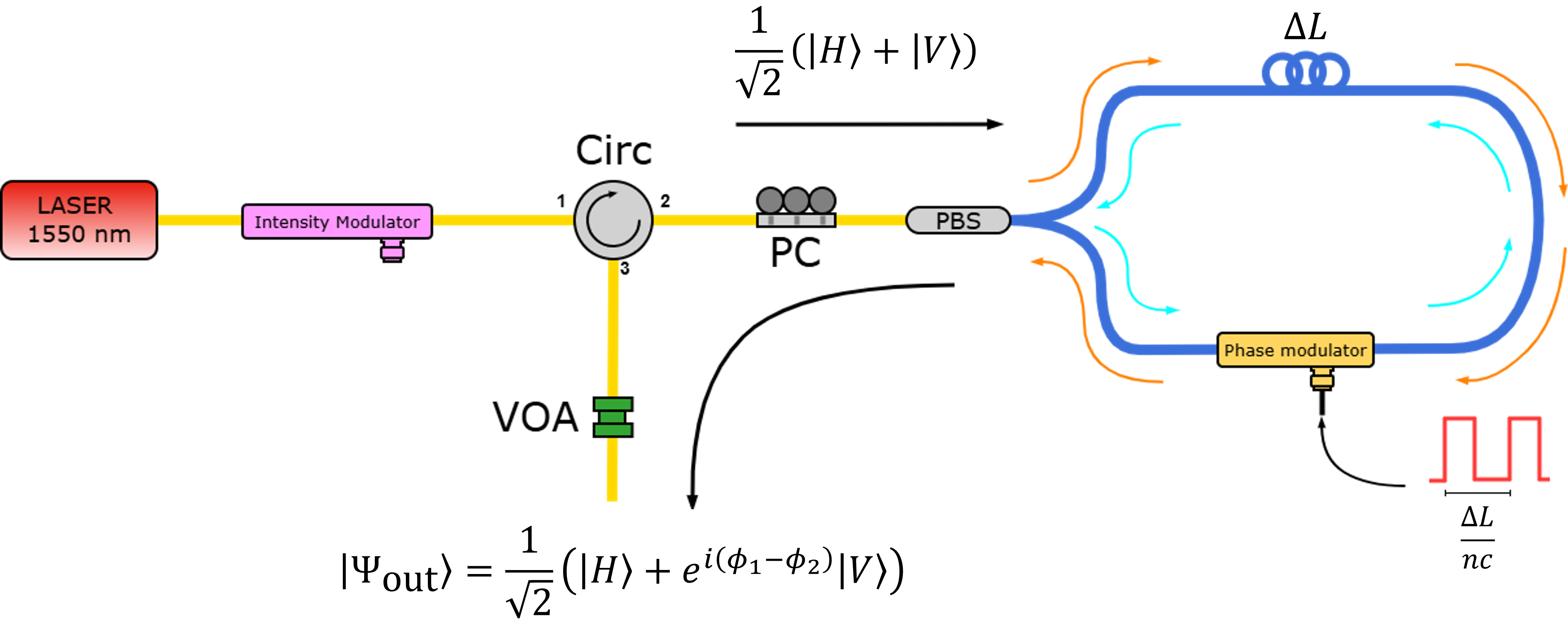}
    \caption{QKD transmitter setup based iPOGNAC-based polarization modulation \cite{agnesi2019}.}
    \label{fig:POGNAC}
\end{figure}

The QKD protocol implemented is the polarization-based efficient-BB84 presented by Grunenfelder et al.~\cite{grunenfelder2018}, where the polarization states are generated using a Sagnac loop based polarization modulator, called iPOGNAC \cite{agnesi2019} depicted in Figure \ref{fig:POGNAC}. The efficient-BB84 QKD is a 3-state protocol, where two states are prepared in the encoding basis and one in the monitoring basis. The encoding basis is used to generate a shared secret key between Alice and Bob, whereas the monitoring basis is used to detect the presence of an eavesdropper. Additionally, to prevent photon-number splitting attacks \cite{huttner1995quantum}, decoy states with a different mean photon number $\nu$ than the signal states with the corresponding mean photon number of $\mu$ are randomly chosen to be transmitted. We choose the encoding basis to be the $X$ basis, consisting of the diagonal and anti-diagonal polarization states, $|D\rangle = \frac{1}{\sqrt{2}}\left( |H\rangle+|V\rangle \right)$ or $|A\rangle = \frac{1}{\sqrt{2}}\left( |H\rangle-|V\rangle \right)$, respectively. Whereas the monitoring basis in our setup is the $Y$ basis, where only right circular-polarized state $|R\rangle = \frac{1}{\sqrt{2}}\left( |H\rangle+i|V\rangle \right)$ is produced. We use the 1-decoy approach, since it is proven to outperform the 2-decoy method for practical experimental settings \cite{rusca2018}. The associated probabilities for producing the signal and the decoy states are denoted as $p_\mu$ and $p_\nu$, respectively.

The probabilities associated to the preparation of polarized states in the encoding and the monitoring bases are denoted by $p^{\text{prep.}}_X$ and $ p^{\text{prep.}}_Y$, respectively. The three polarization states are produced using a phase modulator inside a polarization-maintaining Sagnac loop, such that clockwise and counter-clockwise pulses are modulated differently to generate a desired relative phase \cite{agnesi2019}. This is achieved using a circulator (C), where linearly polarized pulses enter C at port 1 and exit at port 2 into a manual polarization controller (MPC), which transforms the polarization state into the balanced superposition, $|\psi\rangle = \frac{1}{\sqrt{2}}\left( |H\rangle + e^{i\phi_0} |V\rangle \right)$. The light is then split into orthogonal linear polarizations by a fibered polarization beam splitter (PBS), with each output aligned to the slow axis of a polarization-maintaining (PM) fiber, thereby mapping the polarization degree of freedom onto the photon's optical path. The Sagnac interferometer originates at the PBS and is constructed entirely with PM fibers. Within the loop, the vertically polarized component propagates in the clockwise (CW) direction, while the horizontally polarized component propagates counter-clockwise (CCW). The CW component first encounters the $\text{LiNbO}_3$ phase modulator, which introduces a phase $\phi_1$ to the propagating light, and then traverses a PM fiber delay line of length $\Delta L$ before returning to the PBS. In the reverse direction, the CCW component first passes through the delay line and then through the phase modulator, acquiring a phase $\phi_2$. Since both components propagate along the fast axis of the PM fiber, only a single polarization mode is supported in the modulator, eliminating polarization mode dispersion. This guarantees that both components exit the Sagnac loop simultaneously and recombine into a well-defined polarization state at the PBS. The general state emerging from port 3 of the circulator is,
\begin{equation}
    |\psi_{\text{out}}^{\phi_1, \phi_2}\rangle = \frac{1}{\sqrt{2}}\left( |H\rangle + e^{i\left(\phi_1 - \phi_2 - \phi_0\right)} |V\rangle \right).
    \label{eq:pognac}
\end{equation}

Before the light pulse being transmitted to the receiver, the pulses pass through a Variable Optical Attenuator (VOA) that reduces the mean photon number to approximately one photon per pulse. Since photon emission is a statistical process, the photon number per pulse follows a Poisson distribution. The CW pulse reaches the modulator crystal earlier than the CCW pulse by $\frac{\Delta L}{nc}$ seconds, where $\Delta L$ is the length of fiber delay, $n$ is the PM fiber index of refraction and $c$ is the speed of light in vacuum. By carefully timing the voltage applied to the modulator, the output polarization state in Eq. \ref{eq:pognac} can be fully controlled.

For simplicity, setting $\phi_0 = 0$, the three BB84 polarization states can be generated as follows. When no voltage is applied to either pulse, the output is the diagonal state $|D\rangle$,
\begin{equation}
    |\psi_{\text{out}}^{0, 0}\rangle = |D\rangle = \frac{1}{\sqrt{2}}\left( |H\rangle + |V\rangle \right).
    \label{eq:D}
\end{equation}
Applying $V_\pi$ to the CCW component while leaving the CW unmodulated yields the anti-diagonal state $|A\rangle$,
\begin{equation}
    |\psi_{\text{out}}^{0, \pi}\rangle = |A\rangle = \frac{1}{\sqrt{2}}\left( |H\rangle - |V\rangle \right).
    \label{eq:A}
\end{equation}
Finally, applying $V_{\frac{\pi}{2}}$ to the CCW component produces the right-circular state $|R\rangle$,
\begin{equation}
    |\psi_{\text{out}}^{0, \frac{\pi}{2}}\rangle = |R\rangle = \frac{1}{\sqrt{2}}\left( |H\rangle + i |V\rangle \right).
    \label{eq:R}
\end{equation}
More generally, when $\phi_0 \ne 0$, the system can generate any pair of mutually-unbiased bases states lying on the equator of the Bloch sphere.

The generated states are then attenuated to ensure transmission in the single-photon regime before sending them to the receiver over the quantum channel. The QKD receiver setup, described in more detail in the next section, consists of single-photon detectors based measurements, where the measurement bases can be chosen in an active or passive way with bases choice probabilities denoted as $p^{\text{meas.}}_X$ and $p^{\text{meas.}}_Y$. In the 3-state BB84 protocol, while the receiver can project the incoming photons to both states in the encoding basis, only one projection is allowed in the monitoring basis. Therefore, Alice sends the states $\{ |D\rangle, |A\rangle, |R\rangle \}$, whereas Bob projects onto the states $\{ |D\rangle, |A\rangle, |L\rangle \}$, as shown in Figure \ref{fig:field-trial-expt}.

Basis reconciliation step involves Alice disclosing the preparation bases as well as the decoy levels. From the events where Bob's bases choice matches with Alice's, $X$ basis events contribute to the raw key. The remaining events from the $X$ basis and the monitoring $Y$ basis are used to estimate an upper bound on the phase error in the $X$ basis. Error correction is applied on the raw key bits so as to arrive at the same symmetric key except with a negligible probability $\epsilon_c$, whereas privacy amplification procedure bounds the eavesdropper's information on the final secret key through the security parameter $\epsilon_s$.

The security of the efficient-BB84 protocol for the 1-decoy method is analyzed in the finite-key regime \cite{rusca2018} with the secret key length, $\ell$, given as
\begin{equation}
    \ell \leq s_{X,0} + s_{X,1} [1 - h(\phi_X)] - \lambda_{\text{EC}} - 6 \text{log}_2 \left(\dfrac{19}{\epsilon_s}\right) - \text{log}_2 \left(\dfrac{2}{\epsilon_c}\right),
    \label{eq:length}
\end{equation}
where $s_{X,0}$ and $s_{X,1}$ are lower bounds on the number of vacuum and single-photon events in the $X$ basis, respectively, whereas $\phi_X$ is the upper bound on the phase error rate and $h(\cdot)$ denotes the binary entropy. $\lambda_{\text{EC}} = f_{\text{EC}} \cdot n_X^{\text{EC}} \cdot h(Q_{X})$ is the number of bits revealed during the error-correction stage for a raw bits block size $n_X^{\text{EC}}$ (in the experiment, $\lambda_{\text{EC}}$ is the number of disclosed parity bits), error-correction efficiency $f_{\text{EC}}$ and the error rate in the $X$ basis $Q_{X}$. Lastly, $\epsilon_s$ and $\epsilon_c$ are the security and correctness parameters, respectively, defined above. SKR can then be calculated by dividing $\ell$ by the total duration of transmission.

The quantities in Eq.~\eqref{eq:length} are not measured directly, but bounded from the observed detection statistics with the decoy-state method. Let $n_{X,k}$ and $m_{X,k}$ be the number of detections and errors in the encoding basis for intensity $k \in \{\mu, \nu\}$, with $n_X^{\text{EC}} = \sum_k n_{X,k}$, and let $\tau_n = \sum_k p_k e^{-k} k^n / n!$ be the probability that Alice sends an $n$-photon state. Finite-size fluctuations are accounted for with Hoeffding's inequality, through the corrected counts
\begin{equation}
    n^{\pm}_{X,k} = \frac{e^{k}}{p_k}\left[ n_{X,k} \pm \sqrt{\frac{n_X^{\text{EC}}}{2}\ln\frac{19}{\epsilon_s}} \right].
\end{equation}
Following~\cite{rusca2018}, the number of vacuum and single-photon events in the encoding basis are bounded by
\begin{align}
    s_{X,0} &\geq \frac{\tau_0}{\mu - \nu}\left( \mu\, n^{-}_{X,\nu} - \nu\, n^{+}_{X,\mu} \right), \\
    s_{X,1} &\geq \frac{\mu\, \tau_1}{\nu(\mu - \nu)} \left[ n^{-}_{X,\nu} - \frac{\nu^2}{\mu^2}\, n^{+}_{X,\mu} - \frac{\mu^2 - \nu^2}{\mu^2}\, \frac{s^{u}_{X,0}}{\tau_0} \right],
\end{align}
where $s^{u}_{X,0} = 2\left[ \tau_0 \frac{e^{\mu}}{p_\mu}\left( m_{X,\mu} + \delta(m_X) \right) + \delta(n_X^{\text{EC}}) \right]$ is an upper bound on the vacuum contribution, where $\delta(x)=\sqrt{\frac{x}{2}\ln \frac{19}{\epsilon_s}}$. The phase error rate $\phi_X$ cannot be measured in the encoding basis itself and is instead estimated from the errors in the monitoring basis. The same decoy-state analysis applied to the $Y$ basis gives the number of single-photon events $s_{Y,1}$ and an upper bound on the number of single-photon errors,
\begin{equation}
    v_{Y,1} \leq \frac{\tau_1}{\mu - \nu}\left( m^{+}_{Y,\mu} - m^{-}_{Y,\nu} \right),
\end{equation}
from which the phase error rate is bounded by
\begin{equation}
    \phi_X \leq \frac{v_{Y,1}}{s_{Y,1}} + \gamma\!\left( \epsilon_s, \frac{v_{Y,1}}{s_{Y,1}}, s_{X,1}, s_{Y,1} \right),
\end{equation}
where $\gamma(a,b,c,d) = \sqrt{\frac{(c+d)(1-b)b}{c\,d\,\ln 2}\log_2\!\left( \frac{c+d}{c\,d\,(1-b)b}\frac{19^2}{a^2} \right)}$ accounts for the random sampling of the single-photon events between the two bases. The estimate of $\phi_X$ therefore relies on the number of errors $m_{Y,k} = Q_{Y}^k\, n_{Y,k}$ in the monitoring basis, which requires the knowledge of the error rate $Q_{Y}^k$ for each intensity.

However, $Q_Y$ cannot be measured directly in the 3-state protocol, but instead estimated from the detection statistics of the three prepared states \cite{rusca2018security, grunenfelder2018}. For each intensity level $k \in \{\mu, \nu\}$, let $n_{X,k}$ be the number of sifted events in the encoding basis. With respect to cross-basis events, we denote by $n_{RL}$ the number of events in which Alice sends $|R\rangle$ and Bob's monitoring detector clicks, corresponding to state $|L\rangle$; $n_{XL}$ as the number of events in which Alice sends an encoding state ($|D\rangle$ or $|A\rangle$) and the monitoring detector clicks; and $n_{RX}$ the number of events in which Alice sends $|R\rangle$ and Bob detects it in the encoding basis. Normalizing each count by the probability of the corresponding preparation and measurement choices, the error rate in the monitoring basis is
\begin{equation}
    Q_Y = \frac{p^{\text{prep.}}_X p^{\text{meas.}}_X}{2\, n_X^{\text{EC}}}
    \left[ \frac{2\, n_{RL}}{p^{\text{prep.}}_Y p^{\text{meas.}}_Y}
         - \frac{2\, n_{XL}}{p^{\text{prep.}}_X p^{\text{meas.}}_Y}
         - \frac{n_{RX}}{p^{\text{prep.}}_Y p^{\text{meas.}}_X}
         + \frac{2\, n_X^{\text{EC}}}{p^{\text{prep.}}_X p^{\text{meas.}}_X} \right].
    \label{eq:qber_y}
\end{equation}
This expression can be rewritten by defining the normalised encoding-basis yield $\hat{n}_X^{\text{EC}} = n_X^{\text{EC}} / (p^{\text{prep.}}_X p^{\text{meas.}}_X)$ and the following three ratios 
\begin{equation}
    Q_{RL} = \frac{n_{RL}}{p^{\text{prep.}}_Y p^{\text{meas.}}_Y\, \hat{n}_X^{\text{EC}}}, \qquad
    r_1 = \frac{n_{RX}}{p^{\text{prep.}}_Y p^{\text{meas.}}_X\, \hat{n}_X^{\text{EC}}}, \qquad
    r_2 = \frac{2\, n_{XL}}{p^{\text{prep.}}_X p^{\text{meas.}}_Y\, \hat{n}_X^{\text{EC}}},
    \label{eq:ratios}
\end{equation}
to have
\begin{equation}
    Q_Y = Q_{RL} + 1 - \frac{r_1 + r_2}{2}.
    \label{eq:qber_y_ratios}
\end{equation}
Here $Q_{RL}$ is the rate at which $|R\rangle$ is detected as $|L\rangle$. The ratio $r_1$ compares the yield of the monitoring state in the encoding basis detectors with that of the encoding states, and is therefore set by the ratio between the mean photon numbers of the monitoring and the encoding basis states at Alice's output. The ratio $r_2$ compares the probability that an encoding state triggers the monitoring detector, which is $1/2$ for an unbiased measurement (hence, the factor of 2), with its detection probability in the encoding basis. Therefore, $r_2$, measures the relative efficiency of the monitoring and encoding arms of the receiver. For an ideal source and receiver $r_1 = r_2 = 1$ and $Q_Y = Q_{RL}$, and an imbalance in the prepared intensities and/or detection efficiencies shifts $Q_Y$ away from $Q_{RL}$. 

\section{Field-trial demonstration}
\label{sec:field-trial}

\begin{figure}[ht]
    \centering
    \includegraphics[width=0.85\textwidth]{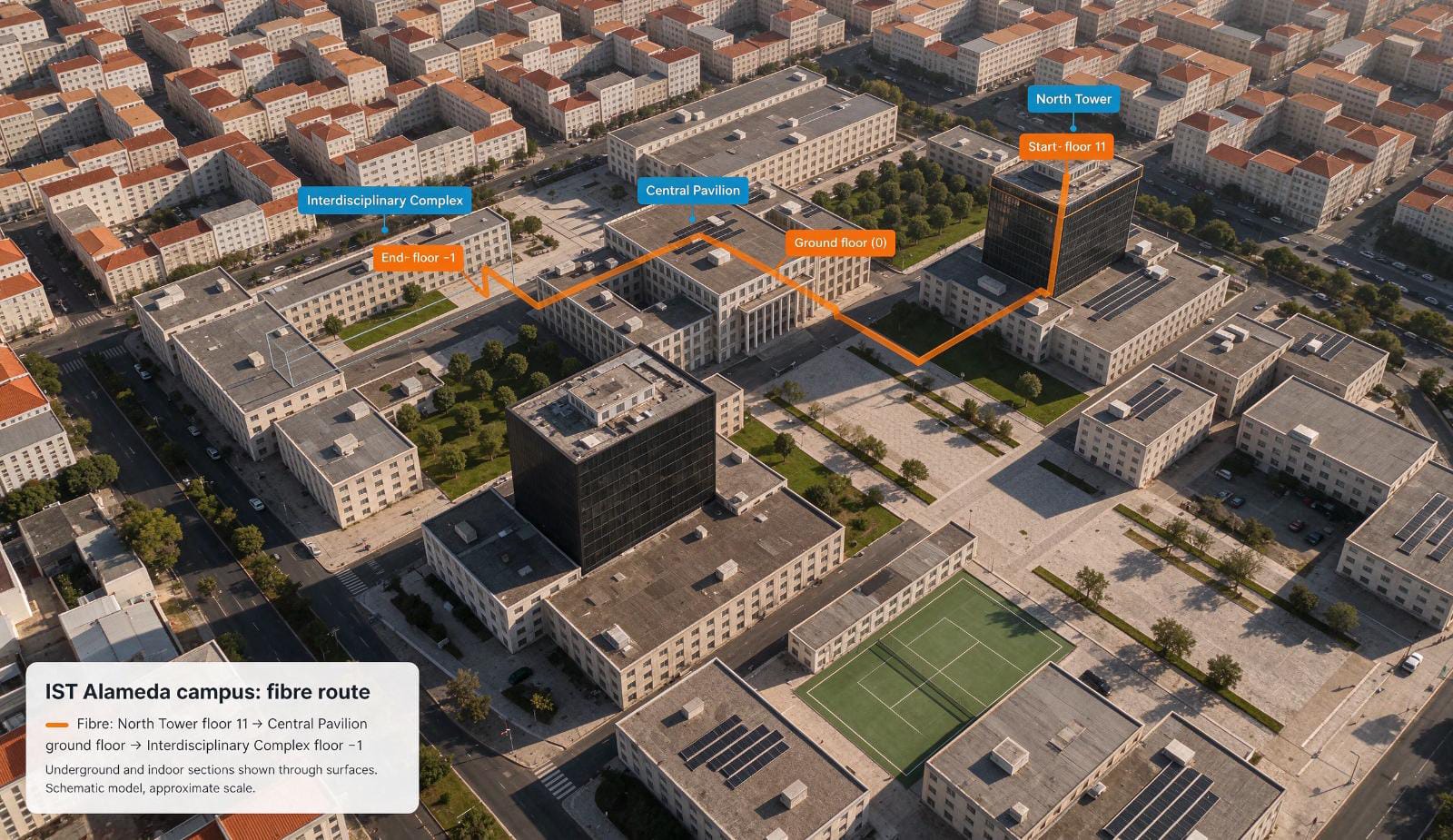}
    \caption{Schematic 3D model of the IST Alameda campus (University of Lisbon) showing the route of the optical fibre link used in this work (orange). The fibre starts on floor 11 (Quantum Photonics Laboratory) of the North Tower and runs down the tower's central service shaft. It then follows a buried duct to the Central Pavilion, crosses its ground floor, and continues underground to floor $-1$ of the Interdisciplinary Complex (Magneto-Optical Trap Laboratory). Links of increasing length were obtained by looping the fibre over 1--4 round trips between the two laboratories, giving total losses of 4.76, 10.21, 16.68 and 24.82~dB, respectively.}
    \label{fig:optical_route}
\end{figure}

The experimental field-trial is demonstrated on the deployed fibers connecting two buildings of the University of Lisbon campus (Instituto Superior Técnico), depicted in Figure \ref{fig:optical_route}. Although Alice and Bob are located at the same laboratory, a set of deployed fibers connecting another laboratory in the campus are used as the communication channel for the signal to go back and forth to use the maximum channel length available. The deployed fiber includes various connections, which resulted in considerably more losses than what is expected for the corresponding length. For instance, for the maximum fiber length of 5.6 km, the measured link loss is around 25 dB. The long term stability of the QKD system is demonstrated for a link loss of around 16 dB, whereas the performance is also evaluated for a total of 5 different channel losses. All optical loss values in the link were obtained using an optical time-domain reflectometer (OTDR).

\subsection{Experimental setup}
\label{subsec:expt}

\begin{figure}[ht]
    \centering
    \includegraphics[width=0.95\textwidth]{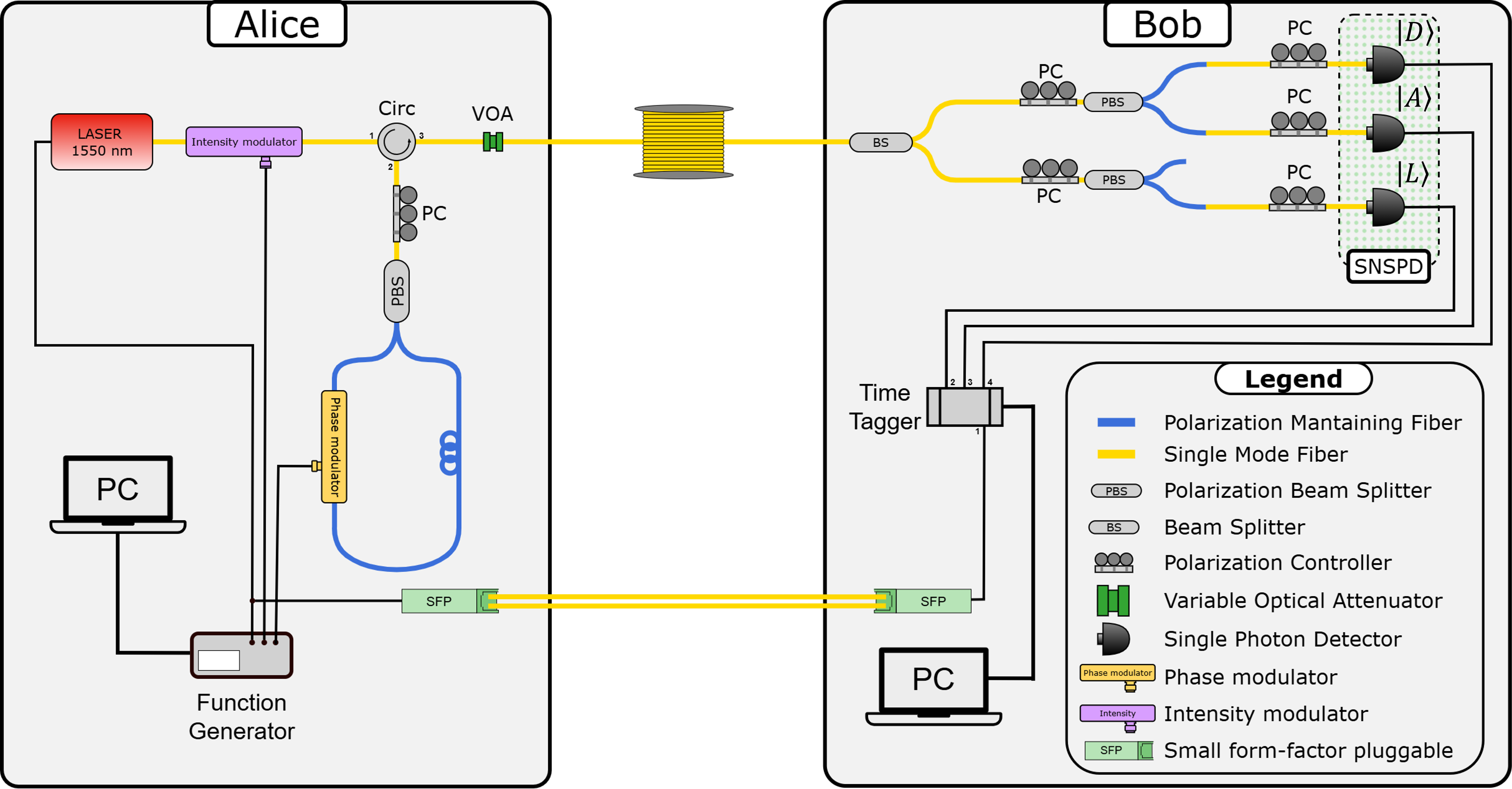}
    \caption{Experimental setup of the efficient-BB84 QKD system. Alice: laser source at 1550\,nm, pulsed at 15\,MHz, followed by an intensity modulator (IM) to the signal ($\mu$) and decoy ($\nu$) levels, followed by the iPOGNAC polarization modulator to prepare $|D\rangle$, $|A\rangle$ and $|R\rangle$. Both modulators are driven by an arbitrary waveform generator (AWG). A variable optical attenuator (VOA) sets the mean photon number at Alice's output. The quantum channel is a deployed fiber across the campus, with the loss set by the number of round trips. Bob: a 50:50 beam splitter performs the passive basis choice, followed by polarization analysis in the encoding ($X$) and monitoring ($Y$) bases and by superconducting nanowire single-photon detectors (SNSPDs, efficiency $\sim$ 90\,\%, jitter at most 50\,ps). Detection events are recorded by a time tagger together with the clock reference. The classical channel uses small form-factor pluggable (SFP) transceivers over a duplex fiber cable to transmit the clock signal reference as well as the classical communication during public discussion.}
    \label{fig:field-trial-expt}
\end{figure}

The experimental setup is depicted in Figure \ref{fig:field-trial-expt}. The transmitter setup consists of a gain-switched laser source (AeroDiode, CCS-1550LD) at 1550 nm light pulses at the rate of 15 MHz with a 5 ns FWHM duration. The light pulses are sent to an x-cut $\text{LiNbO}_3$ intensity modulator (Thorlabs, LNA2322) to generate the signal and decoy states with intensity levels with associated probabilities, $(\mu,p_\mu)$ and $(\nu,p_\nu)$, respectively. The intensity modulated pulses are then coupled into a single-mode (SM) fiber and directed through a circulator (C) that forwards the pulses into the iPOGNAC polarization system described previously in Section \ref{sec:efficient_Pognac}. A MPC ensures the input linear polarization entering the polarization-maintaining Sagnac loop. Inside the polarization-maintaining Sagnac loop, an x-cut $\text{LiNbO}_3$ phase modulator (Exail, MPX-LN-0.1) is asymmetrically placed with an added delay of $\Delta L =5$ m in the CCW direction. This configuration, then, allows the stable generation of the encoding ($X$ basis) and monitoring ($Y$ basis) states with respective probabilities of $p^{\text{prep.}}_X$ and $p^{\text{prep.}}_Y$.

The RF signals used to drive the intensity modulator, phase modulator and the laser are generated by an arbitrary waveform generator (AWG) (RIGOL, DG5000 Pro). The signals to drive the laser and the intensity modulator are timed to incorporate the respective delay experienced by the signal. On the other hand, to drive the phase modulator, the AWG produces two synchronized outputs with an adjustable relative delay. This synchronization ensures that the driving signal is precisely timed to the arrival of each light pulse at the modulator, enabling independent modulation of the CW and CCW components. The three polarization states are thus generated by selectively applying voltages $\{0, V_{\pi}, V_{\pi/2}\}$ to either propagation direction. Note that the AWG limits the state preparation rate due to the rise and fall time of the signals ($\sim 3-4$ ns). In future implementations, the AWG can be replaced by an FPGA paired with an RF amplifier to improve synchronization precision and increase the quantum state generation rate.

The modulated BB84 states are attenuated to single-photon level using a variable optical attenuator (VOA) (Thorlabs, EVOA1550A) before they are transmitted to Bob through the deployed fiber channel. Bob performs the bases selection in a passive way, using a 50:50 beam splitter (BS), giving each incoming photon an equal probability of being directed to either the $X$ basis or $Y$ basis measurement arm, i.e., $p^{\text{meas.}}_X = 1/2 = p^{\text{meas.}}_Y$. A MPC and a fiber PBS in each arm aligned to make projective measurements on the states $|D\rangle$, $|A\rangle$, and $|L\rangle$. Note that the two-detector setup of the simplified-BB84 scheme is not used in this work. Instead, we utilize three super-conducting nanowire single-photon detectors (SNSPDs) (IDQuantique, ID281) with an efficiency up to 90\%, a dark-count rate of around 60 Hz and timing jitters are at most 50 ps. The SNSPD arms also use MPCs to optimize for detection efficiency, since the SNSPD detection efficiency is polarization dependent. The SNSPDs are connected to a time tagger (Swabian Instruments, Time Tagger 20). The resulting detection events are recorded and analyzed on a computer. Synchronization between Alice's and Bob's devices is achieved over a separate fiber using two standard commercial small-form pluggable (SFP) transceiver modules (FS, SFP1G-EX-31-I) and media converters (FS, UMC-GA1F1T), a pair of each connected to Alice's and Bob's computers. This channel is also used to send authenticated classical information during the post-processing stage described below. This is achieved via the use of pre-shared symmetric keys stored at each of Alice's and Bob's computers to generate authentication tags for each communicated public message. We discuss the distribution of these shared symmetric keys in Section \ref{subsec:PQC}. Note that the experimental parameters corresponding to the preparation bases probabilities as well as the signal and decoy state intensities and probabilities are optimized over the secret key rate for each channel loss value.

\subsection{Classical post-processing}
\label{subsec:classical_post}

After Bob communicates the rounds where his detectors clicked, Alice announces the encoding and monitoring bases in the basis reconciliation step. For each basis, she also identifies the signal and decoy rounds. Furthermore, for the $X$ basis rounds for which Bob measured in the $Y$ basis or that correspond to decoy states, Alice also announces the exact state that she prepared.

The error correction algorithm used is the standard Cascade error correction implemented in Python \cite{brunocascade}. The algorithm is applied on blocks of $n_X^{\text{EC}}$ bit long sifted key. After $k=n_X/n_X^{\text{EC}}$ rounds of error correction, privacy amplification is applied on a block of size $n_X$. Toeplitz matrix-hashing based randomness extractor is used for privacy amplification. In here, a $k \times n_X$ sized Toeplitz matrix, $T_{k \times n_X}$, generated using a $(n_X + k -1)$-bit random seed, is applied to the $n_X$-bit long raw key generated post the error correction stage. Toeplitz matrix-based privacy amplification is often used in QKD, due to its advantages associated to using less randomness than a random binary matrix as well as the multiplication speed-up using fast Fourier transform (FFT) \cite{liu2016fit}. While this work does not focus on optimizations to speed-up this multiplication, a modified-Toeplitz matrix by concatenating a $k \times (n_X-k)$-Toeplitz matrix and an $k \times k$-identity matrix, $[T_{k \times (n_X - k)} | I_{k \times k}]$ is used to further reduce the amount of randomness used to $(n_X-1)$ \cite{hayashi2011exponential}. The Toeplitz matrix based randomness extractor is implemented in Python using a seed generated from the Numpy random function, and the generated matrix is stored at each of Alice's and Bob's computers.

\subsection{PQC-based bootstrapping authentication key distribution}
\label{subsec:PQC}

An authenticated classical channel is essential for securing the public communication exchanged during QKD. Commercial QKD systems commonly rely on pre-shared symmetric keys stored at the transmitter and receiver modules to authenticate classical messages. However, secure distribution and management of these keys introduces scalability challenges, particularly in large-scale QKD networks. On the other hand, public-key cryptography based solutions, supported by a public key infrastructure (PKI), offer a more scalable alternative for authentication. In this context, PQC digital signature algorithms, such as the NIST-standardized ML-DSA and SLH-DSA, have been used to provide authentication believed to be resistant to quantum attacks. Although PQC schemes offer computational security, as opposed to the information-theoretic security of symmetric-key based message authentication codes (MACs), their short-term security suffices for the initial authentication phase of QKD \cite{mosca2013quantum}. That is, as long as the PQC schemes used for authentication remain secure until the QKD key generation (post the privacy amplification phase), the forward secrecy of the generated key is guaranteed. Moreover, upon the successful completion of the first round of QKD, a portion of the generated key can be reserved for authenticating the subsequent QKD rounds. With an information-theoretic secure encryption scheme such as one time pad, and the Wegman Carter authentication, one can completely remove the need for a computationally secure system from that point onward. Note that it is required that the length of the generated QKD key is strictly higher than the expected amount of authentication keys to be used in the subsequent QKD round. Therefore, this approach of incorporating PQC based schemes provides practical means of bootstrapping the initial symmetric authentication keys required for QKD. Once the initial QKD round has been successfully completed, the system can solely rely on information-theoretically secure authentication using the generated QKD keys. Consequently, the long-term security of the system relies on the PQC-based authentication remaining unbroken only until the establishment of the first QKD-generated secret key, after which the authentication process can be sustained entirely through information-theoretically secure mechanisms, even if the employed PQC scheme is found to be insecure in the future.

In this work, instead of directly authenticating the classical communication using digital signatures as it has been widely done in the literature \cite{wang2021experimental, garms2024experimental}, we employ an authenticated key establishment procedure.
Our approach is motivated by recent developments in KEM-based authentication in large-scale networks, such as Muckle\# \cite{battarbee2025quantum}. 
For this, we use a post-quantum KEM with ephemeral keys (that are subsequently destroyed) to establish an initial shared secret between Alice and Bob.
This key establishment protocol is authenticated using a post-quantum digital signature scheme whose keys are authenticated by the assumed PKI.
This has two main operational advantages: first, it allows for obtaining a bootstrapping shared secret, which may then be used in a MAC, e.g., the information-theoretic secure Wegman Carter scheme \cite{wegman1981new}; second, if for some reason the shared key is lost, this process can be repeated to again bootstrap the QKD system with a new shared key.
Regarding security, the main advantage is that, assuming that the computational PQC primitives remain secure during a short fixed period that is strictly contained in time, then the whole scheme becomes secure even if these schemes turn out to be broken in the future, providing the desired property of perfect forward secrecy.

Concretely, in our PQC-QKD hybrid scenario, we employ the Wegman Carter authentication scheme \cite{wegman1981new}.
To authenticate the $i$-th classical message $m_i$, a keyed universal hash function $h_{k_h}$ is employed to compute the authentication tag $t_i = h_{k_h}(m_i) \oplus k_o^i$, where $k_h$ determines the hash function and $k_o^i$ is a one-time-pad key. Using key recycling, a fresh one-time-pad key is required for each authenticated message, whereas the same hash key $k_h$ can be reused across multiple authentication rounds \cite{portmann2014key}. 
While the Wegman Carter authentication scheme is able to provide information-theoretic authentication when initialized with truly uniformly random secret keys, the security of the first-round authentication remains only computationally secure due to the previous bootstrapping KEM-based key establishment.
However, since it is strictly contained in a time period during which the PQC schemes are assumed to be secure, the whole scheme becomes information-theoretic secure after that point in time.
Indeed, following the successful completion of the first QKD round (that requires the computational security of the KEM generated key to hold), a portion of the QKD generated secret keys can be reserved to replace the initially KEM-established key.
In the next Section, we also report the estimated time duration for which the computational security of the PQC KEM needs to hold.
Given the short period where the parties must be assumed to be computationally bounded and unable to break the PQC schemes, this procedure allows to completely remove the necessity of pre-shared keys between the parties and enable truly scalable QKD in large networks under weak security assumptions and high computational efficiency.

To demonstrate our setup, we develop a prototype implementation of this classical bootstraping procedure in C, using the libsodium  1.0.22~\cite{libsodium} and liboqs 0.16.0~\cite{liboqs} libraries for the post-quantum KEM and digital signature primitives, respectively.
In particular, for the KEM, we implement a hybrid key establishment via X-Wing~\cite{xwing} (ML-KEM-768 + X25519) from libsodium. 
For a hybrid digital signature scheme, we use a hybrid dual verification of a classical digital signature scheme, Ed25519, via libsodium, and the post-quantum digital signature standard ML-DSA-65 from liboqs.
The execution of the authenticated key establishment works under a PKI that verifies the parties', Alice and Bob, digital signature keys.
The execution goes as follows:
First, Alice generates an ephemeral X-Wing public-private key pair, which she sends to Bob, authenticated with her Ed25519+ML-DSA public key.
Second, Bob encapsulates a shared key using Alice's X-Wing public key, and sends the ciphertext to Alice, authenticated with his Ed25519+ML-DSA public key, and obtains the shared pre-secret.
Third, Alice decapsulates the ciphertext with her X-Wing private key, and obtains the shared pre-secret.
Fourth, both parties use the shared pre-secret to derive a full secret using a HMAC-based key derivation function (HKDF) to a fixed size key (8 KB).
Since a single QKD session consumes approximately $1.75\times10^{5}$ bits ($\approx 22$\,KB) for authentication (Section~\ref{sec:results}), the key establishment is executed three times to obtain a 24\,KB bootstrapping key, sufficient to authenticate the first QKD session.

\section{Results}
\label{sec:results}

The system performance is evaluated in terms of the QBER and SKR as a function of channel losses, while its long-term stability is assessed by monitoring the QBER in the encoding and monitoring bases, together with the intensities of the signal and decoy states over a period of 18 hours. The system achieves a maximum SKR of \num{269.94 +- 1.24} kbps for the setup with only intrinsic losses corresponding to 1.33 dB, while maintaining a SKR of \num{1.09 +- 0.077} kbps at the maximum channel loss of 24.82 dB. During the 18 hour stability experiment, conducted at a channel loss of 16.68 dB, the average SKR is \num{8.2 +- 0.2} kbps. The average QBERs in the $X$ and $Y$ bases are \num{0.24 +- 0.05}\% and \num{0.17 +- 0.06}\%, respectively, indicating stable operation throughout the measurement period.

\begin{table}[h!]
    \centering
    \caption{Experimental parameters for the long-duration QKD experiment.}
    \begin{tabular}{|c | c | c|} 
        \hline
        Parameter & Symbol & value \\
        \hline\hline
        Repetition rate & $R$ & 15 MHz \\ 
        \hline
        Wavelength & $\lambda$ & 1550 nm \\
        \hline
        Detectors efficiency & $\eta$ & $\sim$ 90\% \\
        \hline
        Dark count rate & $R_{\text{dc}}$& $\sim$ 60 Hz \\
        \hline
        Alice's $X$ basis probability    & $p^{\text{prep.}}_X$       & $0.49$ \\
        \hline
        Bob's $X$ basis probability  & $p^{\text{meas.}}_X$       & $0.5$ \\
        \hline
        Signal intensity (mean photon no.)               & $\mu$       & \num{0.264 +- 0.003} \\
        \hline
        Decoy intensity (mean photon no.)                & $\nu$       & \num{0.096 +- 0.002} \\
        \hline
        Signal probability          & $p_{\mu}$   & $0.28$ \\
        \hline
        Acceptance window & $\Delta t$& 6 ns \\
        \hline
        Error correction block size & $n_X^{\text{EC}}$ & $10^5$ \\
        \hline
        Privacy amplification block size & $n_X$ & $10^6$ \\
        \hline
        Correctness parameter & $\epsilon_{c}$ & $10^{-15}$ \\
        \hline
        Secrecy parameter & $\epsilon_{s}$ & $10^{-9}$ \\
        \hline
    \end{tabular}
    \label{tab:sysparams}
\end{table}

Our analysis uses finite-key setting, with data acquisition continuing until the required number of detection events for the predefined block sizes is accumulated. The error correction and privacy amplification block sizes are set to $n_X^{\text{EC}} = 10^5$ and $n_X = 10^6$, respectively. Accordingly, error correction is performed on blocks of $10^5$ bits, and privacy amplification is applied after ten such blocks have been processed. The correctness and secrecy parameters are set to $\epsilon_{c} = 10^{-15}$ and $\epsilon_{s} = 10^{-9}$, respectively. Following the finite-key setting \cite{rusca2018}, parameter estimation takes into account the finite statistical fluctuations as well as the detector efficiency and the dark count probability. The dark count probability $p_{\text{dc}}$ is the probability that a detector registers a click within an acceptance window, defined by a time interval of width $\Delta t$ centered on the expected photon arrival time after each clock pulse, outside of which detection events are discarded. For a detector with dark count rate $R_{\text{dc}}$, it is given by $p_{\text{dc}} = 1 - e^{-R_{\text{dc}}\Delta t} \approx R_{\text{dc}}\,\Delta t$. The privacy amplification procedure explicitly incorporates the information disclosed during error correction, quantified by the actual number of parity bits exchanged between Alice and Bob during the Cascade reconciliation procedure. In our implementation, the reconciliation efficiency of the employed Cascade algorithm turns out to be around $f_{\text{EC}} \approx$ \num{1.12 +- 0.01}, estimated based on the revealed parity bits $\lambda_{\text{EC}}$, error correction block size $n_X^{\text{EC}}$ and the $X$ basis QBER, $Q_X$. The measured number of parity bits revealed by Alice is between 3400 and 5700 bits per block.

Although Bob's measurement basis probabilities are fixed by the passive choice implemented using a beam splitter, the remaining protocol parameters are optimized for each channel loss. These include Alice's preparation bases probabilities, as well as the intensities and probabilities associated with the signal and decoy states. Table \ref{tab:sysparams} summarizes the experimental parameters used for a channel loss of 16.68 dB, corresponding to the long-duration stability experiment. The secret key length $\ell$ is calculated using Equation \eqref{eq:length}, and the SKR is obtained by dividing $\ell$ by the total time required to accumulate 10 blocks of $n_X^{\text{EC}} = 10^5$ bits.

\begin{figure}[ht]
    \begin{subfigure}[b]{0.48\textwidth}
      \includegraphics[width=\textwidth]{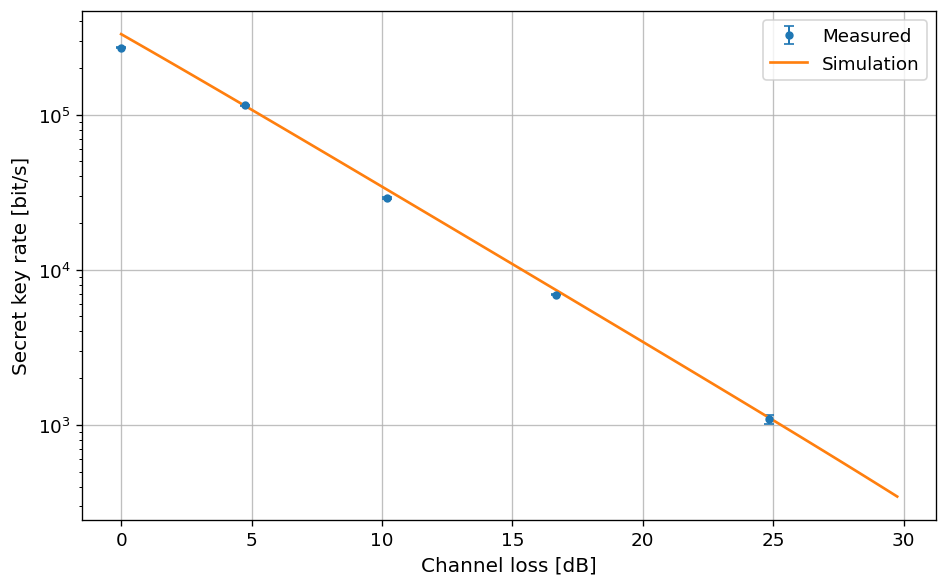}
      \caption{}
      \label{fig:skr}
    \end{subfigure}
    \hfill
    \begin{subfigure}[b]{0.48\textwidth}
      \includegraphics[width=\textwidth]{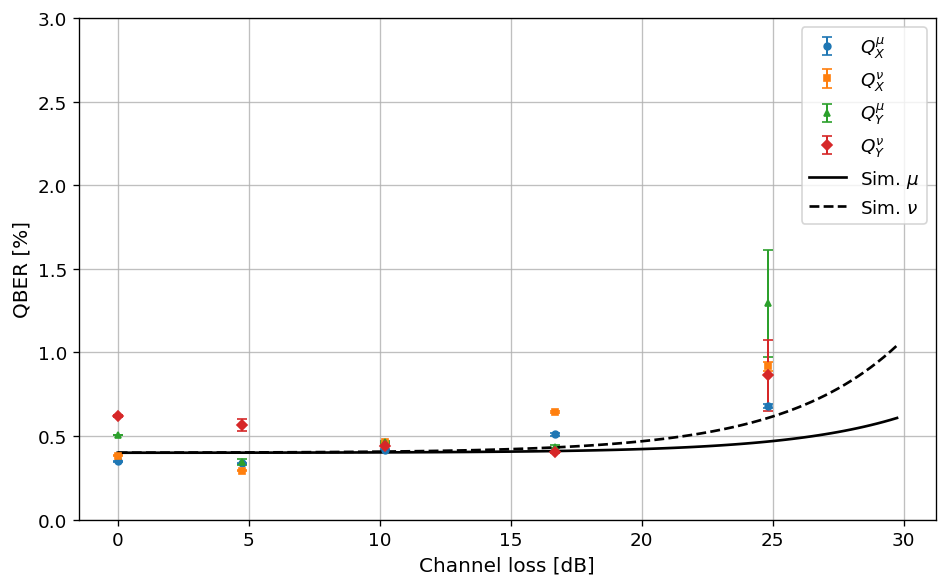}
      \centering
      \caption{}
      \label{fig:qber}
    \end{subfigure}
    \caption{Performance with channel losses: (a) Measured Secret key rate. Each point is the mean over $N=100$ blocks of $n_X^{\text{EC}} = 10^{5}$ sifted bits, and the error bars are the standard error of the mean. The solid line is the simulated SKR including detector efficiency (85–90 \%), dark counts and finite-key effects ($\epsilon_{s} = 10^{-9}$, $\epsilon_{c} = 10^{-15}$). Loss is the measured loss of the deployed campus fiber for each number of round trips. (b) QBERs as a function of the channel loss for the signal (decoy) states in the encoding and monitoring bases, $Q_X^\mu$ and $Q_Y^\mu$ ($Q_X^\nu$ and $Q_Y^\nu$), respectively. Points and error bars are the mean and standard error over the blocks of each acquisition. The solid and dashed lines are the simulated QBER for the signal and decoy intensities, respectively. The increase at high loss comes from dark counts, which become comparable to the detection rate, and affects mostly the decoy state.}
    \label{fig:performance}
  \end{figure}

The system performance is depicted in Figure \ref{fig:performance}. In particular, Figure \ref{fig:skr} presents the experimental and simulated SKR across the investigated channel loss range. The channel loss is varied by changing the number of round trips through the deployed campus fiber, with the loss of each configuration characterized prior to data acquisition. Each data point represents the mean SKR over 100 acquisition blocks, while the error bars indicate the standard error of the mean. The measured SKR decreases by \num{1.12 +- 0.02} dB per dB of channel loss, slightly steeper than the unit slope expected from the linear dependence of the key rate on the channel transmittance for decoy-state BB84~\cite{lo2005decoy, pirandola2017}. This additional decrease, corresponding to a reduction of approximately 23\% per additional dB of loss, originates from the increase of the phase error bound $\phi_X$ and of the relative contribution of dark counts at higher losses, which reduce the fraction of the sifted key that can be distilled into a secret key. The system achieves a maximum SKR of \num{269.94 +- 1.24} kbps at 1.33 dB and maintains a positive key rate of \num{1.09 +- 0.08} kbps at 24.82 dB, the highest loss investigated. This latter value corresponds to approximately 113 km of standard low-loss telecom fiber. At the lowest channel loss, the measured SKR falls $\sim$ 4.6 \% below the simulated prediction. This discrepancy can be partly attributed to the high detection rate at this setting, where the signal-state gain reaches approximately 0.19, corresponding to a total detection rate of around 1.6 MHz. Under these conditions, the SNSPD dead time becomes non-negligible, reducing the effective detection rate. Overall, the experimental SKR closely follows the simulated predictions throughout the investigated channel loss range.

Figure~\ref{fig:qber} shows the QBER in the $X$ and $Y$ bases for both signal and decoy intensities as a function of channel loss. The QBER in both bases remains below 0.7\% across the investigated loss range, except at the highest-loss setting. At this point, the reduced photon detection rate makes detector dark counts increasingly significant, resulting in a noticeable increase in the decoy state QBER. In particular, the $X$ basis QBER remains within the range of 0.34--0.51\% over most of the loss range. This residual error floor is primarily attributed to imperfect polarization alignment, corresponding to an effective polarization misalignment angle of approximately $3$--$4^\circ$. Furthermore, since the polarization controllers at Bob's side are manually adjusted, requiring recalibration for each channel loss configuration, variations in the alignment may also contribute to the observed fluctuations in the QBER. As the channel loss increases, the decoy state QBER begins to rise first. This is expected because the lower decoy state detection probability, approximately $3\times10^{-4}$ per pulse at 24.82 dB, approaches the dark count probability within the detector acceptance window, which is approximately $4 \times 10^{-7}$ per pulse per detector. Consequently, dark counts constitute an increasingly significant fraction of the registered decoy state detections, leading to a higher QBER.

In the three-state protocol, the QBER in the $Y$ basis is not measured directly, but is instead estimated from the detections in the ($X$) basis and the error rates in the monitoring basis~\cite{rusca2018security}. This estimation depends on the calibration parameters $r_1$ and $r_2$ defined in section~\ref{sec:efficient_Pognac}. In particular, $r_1 + r_2 > 2$ underestimates the phase error rate, which makes the estimate non-conservative. For this reason, both ratios are monitored during operation and kept close to unity, by adjusting the intensity of $|R\rangle$ at the transmitter and balancing the detection efficiencies at the receiver. Before each acquisition, the system is calibrated to ensure that $(r_1 + r_2)/2 \leq 1$, thereby ensuring a conservative estimate of the $Y$ basis QBER. With this calibration, the estimated QBER remains within 0.25--0.5\% of the measured $|R/L\rangle$ error rate. Table~\ref{tab:lossparams} lists all the experimentally measured parameters used to calculate the final SKR at each channel loss.

\begin{table}[htbp]
    \centering
    \caption{Experimental results for different channel losses.}
    \label{tab:lossparams}
    \resizebox{\textwidth}{!}{%
    \begin{tabular}{|c|c|c|c|c|c|c|c|}
         \hline
         Loss [dB] & Time per block [s] & $Q_X$ [\%] & $Q_Y$ [\%] & $s_{X,1}/n_X^{\text{EC}}$ & $\phi_X$ [\%] & $f_{\mathrm{EC}}$ & SKR [kbps] \\
         \hline\hline
         1.33  & \num{0.293 +- 0.0001} & \num{0.348 +- 0.003} & \num{0.507 +- 0.002} & \num{0.8415 +- 0.0006} & \num{0.01 +- 0.01} & \num{1.120+-0.0005}& \num{269 +- 2} \\
         \hline
         4.76  & \num{0.589 +- 0.0002} & \num{0.337 +- 0.003} & \num{0.35 +- 0.02} & \num{0.7948 +- 0.0006} & \num{0.33 +- 0.08}  & \num{1.112+-0.0006}& \num{117 +- 1} \\
         \hline
         10.21 & \num{1.918 +- 0.0007} & \num{0.417 +- 0.003} & \num{0.465 +- 0.004} & \num{0.7811 +- 0.0006} & \num{0.50 +- 0.08}  & \num{1.123+-0.0004}& \num{29.1 +- 0.4} \\
         \hline
         16.68 & \num{9.320 +- 0.007}         & \num{0.510 +- 0.007} & \num{0.437 +- 0.008} & \num{0.7619 +- 0.0009} & \num{0.70 +- 0.09}  & \num{1.121+-0.0006}& \num{6.91 +- 0.07} \\
         \hline
         24.82 & \num{41.5 +- 0.1}            & \num{0.68 +- 0.01}   & \num{1.3 +- 0.3}   & \num{0.681 +- 0.002}   & \num{3.8 +- 1.0}      & \num{1.136 +- 0.001}         & \num{1.09 +- 0.08} \\
         \hline
    \end{tabular}}
\end{table}

\begin{figure}[ht]
    \begin{subfigure}[b]{0.48\textwidth}
      \includegraphics[width=\textwidth]{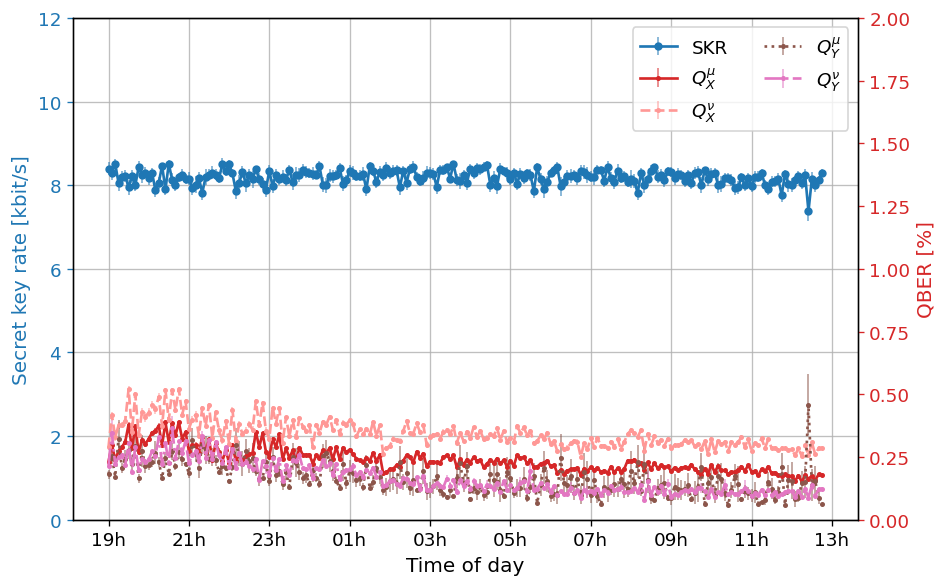}
      \caption{}
      \label{fig:skr_time}
    \end{subfigure}
    \hfill
    \begin{subfigure}[b]{0.48\textwidth}
      \includegraphics[width=\textwidth]{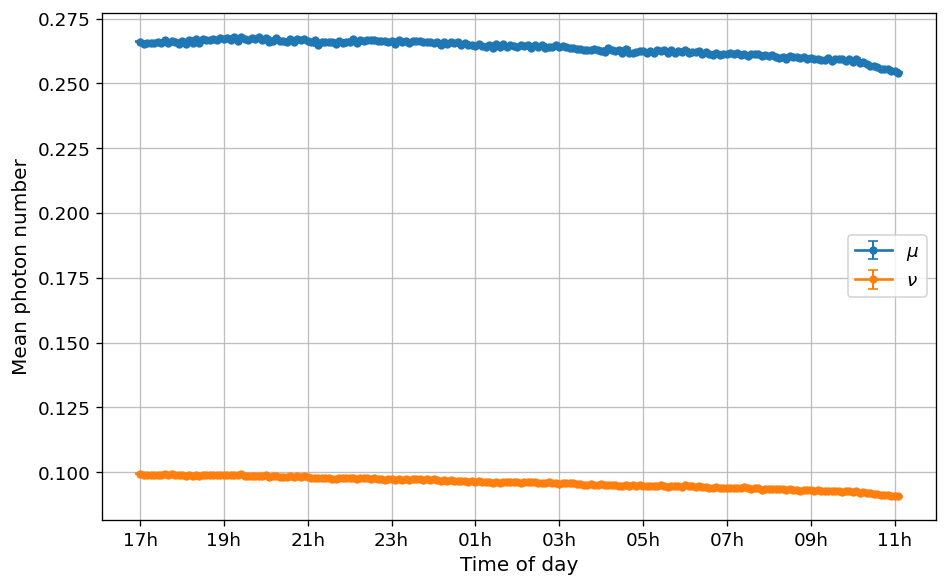}
      \centering
      \caption{}
      \label{fig:mpn_time}
    \end{subfigure}
    \caption{Long-term operation over a channel loss of 16.68\,dB: (a) The secret key rate (left axis) and the QBERs of the signal (decoy) states in the encoding and monitoring bases, $Q_X^\mu$ and $Q_Y^\mu$ ($Q_X^\nu$ and $Q_Y^\nu$), respectively (right axis) shown as a function of the time of day. Each point corresponds to an acquisition started every 5\,min and is the mean over 15--16 blocks of $n_X^{\text{EC}} = 10^5$ bits. The error bars are the standard error of the mean. (b) Mean photon numbers of the signal ($\mu$) and decoy ($\nu$) states shown as a function of the time of day. They are inferred from the vacuum probability of the encoding states measured at Bob and referred to Alice's output by correcting for the channel loss and the passive basis choice. Points and error bars are the mean and standard error of the mean over the blocks of each acquisition.}
    \label{fig:stability}
  \end{figure}

To study its stability, the system ran continuously for 18.2 hours, with calibration performed only at the beginning of the experiment and is then left running without any recalibration, polarization tracking or active stabilization. Each data point of SKR, QBERs and the intensities in Figure \ref{fig:stability} is averaged over a time window of 300 s. The system produced a mean SKR of $8.2 \pm 0.2$\,kbps on the 16.68\,dB channel (3 round trips), with $Q_X = 0.24 \pm 0.05\,\%$ and $Q_Y = 0.17 \pm 0.06\,\%$. Figure~\ref{fig:skr_time} shows the SKR and the QBERs for each time window over the full acquisition.

The key rate remains remarkably stable throughout the experiment, with a relative standard deviation of 2\,\% and a minimum of 7.4\,kbps. Both QBERs decrease slowly over time: $Q_X$ goes from $0.31\,\%$ in the first 3\,hours to $0.19\,\%$ in the last 3\,hours, and $Q_Y$ from $0.24\,\%$ to $0.13\,\%$, which indicates that the polarization reference set at the beginning of the experiment remained well aligned, with no drift requiring recalibration. Fluctuations of the SKR are mainly correlated with $Q_Y$ (correlation coefficient \num{-0.47 +- 0.05}), while no correlation is observed with $Q_X$. These results show that a single initial calibration sustains a stable key generation for more than 18\,hours on this deployed link.

The mean photon numbers of the signal and decoy states, inferred from the vacuum probability at Bob and referred to Alice's output through the measured channel loss, are shown in Figure~\ref{fig:mpn_time}. Both decrease slowly and monotonically over the whole acquisition, with $\mu$ dropping from $0.266$ to $0.257$ ($-3.4\,\%$) and $\nu$ from $0.099$ to $0.092$ ($-7.5\,\%$). Since both intensities decrease together, the dominant contribution is a common attenuation, such as a drift in the laser power, in the channel transmission or in the detection efficiency, which cannot be distinguished from the detection statistics alone. The ratio $\nu/\mu$ decreases from $0.373$ to $0.357$, indicating a small drift of the intensity modulator bias in addition to a common attenuation. A $3.4\,\%$ reduction of $\mu$ lowers the expected SKR by about $2.5\,\%$, consistent with the slight decrease of the SKR observed in the last hours.

To assess the time window over which the computational security of the initial bootstrapping key must be maintained, we measured the processing time of the most computationally demanding procedures, namely error correction and privacy amplification. During a single QKD session, the total runtime of the 10 error-correction rounds is \num{5.2 +- 0.2}\,s ($\sim$ 0.5 s per block of $10^{5}$), while the single privacy amplification round takes \num{0.096 +- 0.001}\,s. Consequently, the PQC KEM-derived keys only need to remain computationally secure for approximately \num{5.3 +- 0.2}\,s, corresponding to the time required to generate the first QKD keys. It is worth noting that this estimate excludes the quantum transmission time required to accumulate a sufficient number of sifted bits for post-processing. This time is not included because it is relevant primarily to real-time implementations and is dependent on the channel losses. We, therefore, obtain an estimate of the computational security window based solely on the post-processing requirements, excluding the quantum channel conditions.

To evaluate the balance between authentication key consumption and generation, we estimated the total number of secret bits required to authenticate the public communication exchanged during each QKD session and compared this with the amount of secret key generated. Following the key-recycling approach for the Wegman Carter authentication scheme \cite{portmann2014key}, a 128-bit key $k_h$ for universal hashing is reused across the public communications during a single QKD session, while a fresh 128-bit one-time-pad key $k_o^i$ is consumed for each authenticated public communication. Based on the parity information messages exchanged during each error correction round, we estimate approximately $1300-1400$ authenticated communication rounds per session. Including the authentication requirements for sifting and privacy amplification rounds, this amounts to a total consumption of \num{1.75 +- 0.10 e5} secret bits per session. In comparison, the generated QKD secret key is approximately $\sim 6.6-8.0 \times 10^5$ bits per session. This demonstrates that the key generation rate comfortably exceeds the authentication key consumption, allowing sufficient fresh secret bits to be allocated for authenticating the subsequent QKD sessions while retaining the remaining keys for encryption.

\section{Conclusions}
\label{sec:conclusions}

In this work, we have experimentally demonstrated a 1550 nm polarization-encoded efficient-BB84 QKD system over deployed optical fiber in a Lisbon university network. The 3-state 1-decoy discrete-variable QKD system employs an iPOGNAC-based transmitter with intensity modulation together with SNSPDs at the receiver. We evaluated the system performance under five different incremental channel losses, ranging from 1.33 to 24.82 dB. The experimental results demonstrate positive secret key generation throughout the investigated loss range, achieving secret key rates from 270 kbps at the lowest loss to 1.09 kbps at the highest loss. Furthermore, the measured key rates closely follow the simulated predictions, confirming the expected performance of the finite-key 1-decoy protocol under different channel conditions.

We also investigated the long-term stability of the system through a continuous acquisition lasting around 18 hours at a channel loss of 16.68 dB. Without active polarization stabilization, the system maintained an average SKR of 8.2 $\pm$ 0.2 kbps, with average QBERs of 0.24 $\pm$ 0.05\% and 0.17 $\pm$ 0.06\% in the encoding and monitoring bases, respectively. Importantly, a positive secret key rate was sustained throughout the experiment, never falling below 7.4 kbps. These results demonstrate the robustness of our QKD implementation under realistic deployed fiber conditions and its suitability for extended QKD operation without continuous active stabilization.

Finally, we presented a PQC-based authentication approach that employs a KEM to bootstrap the initial symmetric authentication keys required for QKD. By combining KEM-based key establishment with Wegman Carter authentication, the proposed approach enables the transition from computationally secure initial authentication to information-theoretically secure authentication using QKD-generated keys in subsequent rounds. This provides a practical alternative to relying on permanently pre-shared symmetric keys or directly authenticating every classical communication message using computationally heavy PQC digital signatures. In particular, the time required to establish the first QKD generated key is reported to be around 5.3 s, which determines the period over which the computational security of the initial PQC-based bootstrap needs to hold.

Overall, this work demonstrates the feasibility of combining stable, finite-key efficient-BB84 QKD over deployed optical fiber with PQC-based authentication bootstrapping. The results contribute towards practical hybrid quantum-safe communication infrastructures, in which PQC provides an initial authentication mechanism while QKD subsequently supplies information-theoretically secure keys for both key generation and classical communication authentication.

\section{Acknowledgments}

This work is funded by national funds through FCT – Fundação para a Ciência e a Tecnologia, I.P., and, when eligible, co-funded by EU funds under project/support UID/50008/2025 – Instituto de Telecomunicações, with DOI identifier https://doi.org/10.54499/UID/50008/2025. The authors thank Capgemini Engineering for providing part of the equipment used in this experiment. The authors thank the support from the European Commission (EC) through the project PTQCI (DIGITAL-2021-QCI-01) and IberianQCI (24-EU-DIG-IberianQCI). G.L.T. acknowledges the support of FCT through scholarship FCT 2024.01028.BD.

\bibliography{bibl.bib}{}
\bibliographystyle{unsrt}

\end{document}